\documentclass[namedreferences,hyperref,optionalrh]{spr-sola}
\usepackage{graphicx}        
\usepackage{color}           
\usepackage{breakurl}                         

\usepackage{ulem}

\usepackage{amsmath}
\usepackage{amssymb}

\usepackage{xcolor}

\usepackage{hyperref}
\usepackage{soul}
\hypersetup{
    colorlinks=True,
    linkcolor=blue,
    filecolor=magenta,
    citecolor=blue,
    urlcolor=blue,
    pdftitle={Overleaf Example},
    pdfpagemode=FullScreen,
}
\usepackage{lscape}

\chardef\us=`\_

\begin{document}


\begin{frontmatter}
\title{Metric Type II Radio Emission associated with Coronal Mass Ejections of Large Angular Widths: Some New Insights}

\author[addressref={aff1},email={521ph1012@nitrkl.ac.in}]{\inits{M.}\fnm{Mayank}~\snm{Rajput}\orcid{0009-0007-0043-6823}}
\author[addressref=aff1,corref,email={bisois@nitrkl.ac.in}]{\inits{S.K.}\fnm{Susanta Kumar}~\snm{Bisoi}\orcid{0000-0002-9448-1794}}
\author[addressref=aff2,email={janardhan.padmanabhan@gmail.com}]{\inits{Janardhan P.}\fnm{Janardhan Padmanabhan}~\snm{}\orcid{0000-0003-2504-2576}}
\address[id=aff1]{Dept. of Physics and Astronomy, National Institute of Technology, Rourkela 769008, India.}
\address[id=aff2]{Astronomy and Astrophysics Division, Physical Research Laboratory, Ahmedabad 380009, India.}

\runningauthor{M Rajput et al.}
\runningtitle{Metric Type II Solar Radio Bursts and Associated CMEs}

\begin{abstract}
Solar Type II radio bursts are manifestations of shocks produced by explosive and eruptive solar activities, such as solar flares and coronal mass ejections (CMEs). Metric range or coronal Type II bursts are interesting because of their association with CME-driven shocks. It is therefore important to examine the correlation between the properties of Type II bursts and the associated CMEs. This, in turn, would be useful in understanding the impact of CMEs on space weather and geomagnetic activity. We conducted a statistical study of {23} coronal Type II radio bursts in the frequency range 150--20 MHz, for the period spanning 2007\,--\,2024. We correlated the frequency bandwidth of coronal Type II bursts with the angular width of the associated CMEs. Our investigation showed {an anti-correlation between the two quantities with a correlation coefficient of $\approx$ -0.62}. We further deduced the height of the Type II bursts (r$_{Type II}$) at the onset time of the burst and compared them with the estimated height of the associated CMEs/shocks (r$_{CME}$). With the exception of one event, for the rest of the Type II burst events, r$_{Type II}$ was $<$ r$_{CME}$. The results suggest that the CMEs with large angular widths produce narrow-band Type II emissions, and these Type II emissions could possibly be produced in the flank region of the CME-driven shock rather than at the shock front.
\end{abstract}
\end{frontmatter}
\section{Introduction} 
Type II bursts are narrow-band radio emissions, and when viewed on a radio dynamic spectra, they drift slowly from high to low frequency {\citep{payne1947relative}}.
{It is widely accepted that Type II bursts are produced by shock waves moving through the heliosphere {\citep{nelson1985type}}, which can accelerate electrons in the solar corona to excite plasma Langmuir waves at the local plasma frequency ($f_p$) and/or its harmonics ($2f_p$).} The conversion of these plasma waves to electromagnetic radiation is observed as Type II bursts {\citep{wild1954harmonics,nelson1985type}}. Shock waves in the corona are generated by various types of explosive solar surface phenomena, among which flare blast waves {\citep{klassen1999radio,vrvsnak2001solar,magdalenic2010origin,nindos2011relationship,alissandrakis2021multiwavelength,morosan2023type,vasanth2024coronal}} and coronal mass ejections (CMEs){\citep{aurass1997coronal,cliver1999origin,maia1999development,lara2003statistical,gopalswamy2006solar,kumari2017new,zhang2024type,vasanth2025observations,vasanth2026shock,nedal2026multiple}} are considered to be their main drivers. Type II radio bursts observed in the outer corona and the interplanetary medium are usually referred to as DH-Type II bursts, while near-sun or coronal Type II radio bursts are typically referred to as metric Type II bursts. Attributing the exact cause of coronal or metric Type II bursts is still debatable because of the close timing of the flare impulsive phase and CME acceleration when a metric Type II burst occurs. Statistical results using white light, X-ray, H$_{\alpha}$, and radio spectral observations indicate that most of the metric type II bursts are produced by CME-driven shocks {\citep{classen2002association,cliver2004coronal,mancuso2004coronal,cho2005examination,cho2008low,gopalswamy2006coronal,kumari2023type}}.

CMEs are bright transient features observed in white light coronagraph images, which expand as they move away from the Sun. These are the result of large-scale plasma and magnetic field eruptions from the solar corona. The CMEs have a huge mass of $10^{11}$-$10^{13}$ kg \citep{gopalswamy1992estimation,chen2011coronal} and travel at very high speeds (500--1000 km/s) through the heliosphere {\citep{gosling1976speeds,gopalswamy2009soho,shen2022propagation}}. CMEs possess a variety of shapes, and a typical CME has a three-part structure: a bright leading edge, a dark cavity, and a bright core. The spatial span of CME in degrees with respect to the Sun center is typically referred to as the angular width of the CME. The CME angular width is usually estimated from the span of the position angles of the CME's outer edges with respect to the Sun center in the coronagraph's sky plane. The CME angular widths for coronagraph images from the Large Angle and Spectroscopic Coronagraph (LASCO) {\citep{brueckner1995large}} onboard the Solar and Heliospheric Observatory (SOHO) {\citep{domingo1995soho}} are obtained using the above method and are listed in the SOHO/LASCO CME catalog{\footnote{\url{https://cdaw.gsfc.nasa.gov/CMElist/}}} {\citep{yashiro2004catalog}}. Several previous studies \citep{gopalswamy2009expansion,compagnino2017statistical,li2023quasiperiodic,dharmashaktu2024statistical,onyemaechi2025statistical,zhou2025hemispheric} have used the CME angular width from the SOHO/LASCO catalog to understand various CME characteristics. These CME angular widths are two-dimensional (2D) sky-plane projected angular width of the CME, which is actually not the real CME angular width in three-dimensional (3D) space, but it represents the CME's apparent angular width. {If a CME erupted close to the limb}, where the three-part structure of the CME is clearly visible, then the 2D angular width estimated is actually close to the real angular width. However, if the CME is away from the limb, then the estimated 2D angular width could be overestimated.  For example, halo CME has a 2D angular width of $\sim$360$^{\circ}$, which does not necessarily represent the real angular width. The large apparent angular width is because of the ejected material being directed along the line of sight (LOS) toward the observer as the halo CME originates close to the central meridian of the Sun. Therefore, to estimate the 3D angular width of CMEs, some 3D reconstruction methods are used. One such 3D reconstruction method is a forward fitting Graduated Cylindrical Shell (GCS) model \citep{thernisien2006modeling,thernisien2009forward}, which employs multi-vantage-point observations of C2 and C3 coronagraphs on board SOHO/LASCO and Sun Earth Connection Coronal and Heliospheric Investigation (SECCHI) {\citep{howard2008sun}} on board the twin spacecrafts Ahead (A) and Behind (B) of Solar Terrestrial Relations Observatory (STEREO) {\citep{2008SSRv..136....5K}}. \cite{zhao2017correlation} correlated the 3D angular width of the CMEs obtained using the GCS model with different characteristics of their source regions. {Several other studies have also used the GCS model to estimate various geometrical parameters of the CMEs, including their 3D angular width. \citep{hu2017multi,zhong2021three,zhuang2024combining,koya2024assessment}.}

When CMEs move through the corona and interplanetary medium, they interact with the surrounding plasma and generate shock waves if their speed is greater than the characteristic speed of the medium, known as the Alfv\'en speed {\citep{gosling1974mass,gosling1976speeds}}. The region ahead of the CME-driven shock is called the upstream region. Similarly, the region behind the shock is called the downstream region. In the case of single-lane (or simple) fundamental Type II bursts, shock drift acceleration of electrons occurs only in the upstream region. The Type II bursts, when viewed with high-resolution spectral observations, exhibit a variety of complex and fine structures, such as split-band {\citep{smerd1974split,smerd1975split,ramesh2022polarization,bhunia2023imaging,zhang2024imaging,normo2025imaging} and multi-lane {\citep{lv2017sources,zucca2025source,morosan2025resolving,nadiya2026multi}} Type II bursts. In the case of split-band Type II bursts, the acceleration of electrons occurs in both the upstream and downstream regions, leading to the generation of the lower-frequency and upper-frequency lanes of split-band Type II emission {\citep{smerd1974split,smerd1975split}}. On the other hand, a multi-lane Type II radio burst consists of three or more parallel lanes, which are the result of simultaneous acceleration of electrons across different parts of the shock front \citep{lv2017sources,koval2023morphology}. The complex features of Type II solar radio bursts provide valuable diagnostics of shock properties and their interaction with the corona's ambient plasma.

Coronal Type II bursts result from the shock-accelerated electrons, which are efficiently accelerated from those locations, where the shock is supercritical and has a quasi-perpendicular geometry, {\it{i.e.}}, the angle between the shock normal and the upstream magnetic field is $>$ $45^{\circ}$ {\citep{zucca2018shock,kouloumvakos2021coronal,morosan2025resolving}}. Such favorable conditions could be met on the front side of CMEs. Thus, it is suggested that the Type II emission is produced in the CME front side.
 In another study {\citep{morosan2025resolving}}, Type II emissions were observed in the southern and northern flanks of the CME shocks, whereas no radio emission was observed in the CME front. It has been suggested that no emission in the CME front side is most likely owing to a quasi-parallel condition in the CME front region. However, they can efficiently produce Type II emissions from their laterally expanding quasi-perpendicular flank regions. In addition to the shock geometry, the shock speed also plays an important role in producing Type II emission {\citep{knock2003theoretically,knock2005type}}. If the Alfv\'en speed is less than the shock speed, then one can observe Type II radio emissions {\citep{hou2023type,cui2026solar}}. In a study, \cite{zucca2014formation} found that Type II radio bursts originated from the flank of the CME, where the Alfvén speed was minimum. \cite{morosan2019multiple} found that the CME was fast enough at the flank to produce a shock wave, eventually resulting in Type II radio emission in the flank of the CME. Several studies suggest that the shock flank can be an efficient source of Type II radio bursts \citep{majumdar2021imaging,morosan2022shock,ramesh2023solar,zucca2025source}. {{Multiple shock waves within a CME can also be sources of a Type II burst. The burst comprises of a herringbone and three parallel lanes {\citep{feng2025multiple}. They found that the source of the herringbone structure was produced from the flank of a CME-driven shock, while the three parallel lanes were produced by the shocks associated with coronal loops within the CME.}}

Since coronal Type II bursts are remote signatures of CME-driven shocks propagating through the heliosphere, their drift rates can be used to predict the arrival time of the shocks and CMEs \citep{li2024predicting,wang2026new}. Therefore, studying coronal Type II bursts is important for understanding space weather effects and issuing warnings. \cite{devi2024type} performed a statistical study of metric to DH Type II radio bursts and the associated space weather phenomena such as CMEs, solar flares, and filament eruptions. Their results showed that radio signatures of shock waves are closely associated with solar eruptions. This further highlights the importance of Type II radio bursts as a reliable indicator of the CME-driven shock's arrival at 1 AU and, hence, for predicting space weather activity. Thus, a study on the relationship between the properties of Type II bursts and CMEs would be significant in space weather forecasting. A study by {\cite{ramesh2022new}} compared the instantaneous frequency bandwidth of coronal Type II bursts, observed with the Gauribidanur LOw-frequency Solar Spectrograph (GLOSS) in the frequency range 85--35 MHz, and the angular extent of the associated CMEs. From a sample of 26 coronal Type II bursts, they reported a significant (71\%) positive correlation between the frequency bandwidth of coronal Type II bursts and the angular width of associated CMEs. In their study {\citep{ramesh2022new}}, they used the angular width of the CMEs from the SOHO/LASCO CME catalog.
We have therefore revisited the correlation between the CME angular widths and the frequency bandwidth of metric Type II bursts. In the present study, we estimated the 3D angular width of CMEs using the GCS model. The estimated angular widths of the CMEs are then correlated with the frequency bandwidths of the Type II bursts to re-examine the relationship between CMEs and Type II bursts. Our study provides new insights into the spatial distribution of electron acceleration regions in shocks associated with CMEs of varying 3D angular widths. Data used and the relevant instrument details are described in Section \ref{obs}. The methodology for estimating the frequency bandwidth and heights of Type II bursts, and the estimation of the 3D angular width and heights of the associated CMEs, are described in Section \ref{analysis}. Results of the study are presented in Section \ref{res}, while, finally, summary and discussion are presented in Section \ref{discuss}.

\section{Data and Instruments}
\label{obs}
\subsection{Data}
We used three Type II burst catalogs, {{comprising a total of 396 events,}} to select the Type II burst events for our study. The three catalogs used were from radio observations made by the following different instruments, 1) The Gauribidanaur LOw-frequency Solar Spectrograph (GLOSS) from the Gauribidanaur observatory\footnote{\url{https://www.iiap.res.in/solarradioimages\#spectrogram}} {\citep{kishore2014gauribidanur}}, operated by Indian Institute of Astrophysics (IIA), Bangalore, India,  2) The Culgoora radio spectrograph$\footnote{\url{https://www.sws.bom.gov.au/World\_Data\_Centre/1/9}}$ {\citep{prestage1994new}}, and 3) the e-CALLISTO \footnote{\url{https://www.e-callisto.org/Data/data.html}} {\citep{benz2009world}}. For these 396 events, we obtained and analyzed the radio dynamic spectrum from the e-CALLISTO spectrograph corresponding to each Type II burst.

To obtain the angular width of the CMEs, we used the SOHO/LASCO CME catalog. The active regions (ARs) associated with the CMEs were obtained from the DST link in the SOHO/CME catalog. For the estimation of angular width using the GCS model, we used multi-vantage-point observations of white-light images of CMEs obtained by C2 and C3 coronagraphs on board SOHO/LASCO and by SECCHI on board STEREO A and B. The data from three instruments of SECCHI, the Extreme Ultraviolet Imager (EUVI), Inner Coronagraph (COR1), and Outer Coronagraph (COR2), were used for the estimation of the angular width of the CMEs.

For estimation of CME heights, we used direct imaging observation of CMEs obtained from the Atmospheric Imaging Assembly (AIA) onboard the Solar Dynamics Observatory (SDO) (SDO/AIA). The field of view (FOV) of the SDO/AIA is $\sim$1.2$R_{\odot}$. Hence, if a CME eruption occurs beyond the FOV of SDO/AIA, we used LASCO/C2 coronagraph images that can observe the CMEs beyond 2R$_\odot$ to estimate the CME heights. Direct images were used from either SDO/AIA and/or LASCO/C2 for estimation of the CME heights, when these are available at the onset time of the Type II burst, and if these are not available, we used the direct images from either SECCHI/EUVI or SECCHI/COR1 onboard STEREO A and B. 

To reconstruct the shock surface associated with the CMEs of 25 October 2013, 20 April 2014, and 08 March 2011, we used AIA 193 {\AA} and EUVI 195 {\AA} observations to fit it.

\begin{table}
 \caption{Type II burst catalogs as prepared by 1) GLOSS,  2) Culgoora,
and 3) e-CALLISTO.}
 \label{tbl:1}
\begin{tabular}{cccccc}     
\hline
 \multicolumn{1}{c}{}
Serial& Instrument& Frequency& Study&    Number & Selected\\
No.&           & Range&     Interval& of Events & Events \\
\hline

1&GLOSS&85--35 MHz &2009--2021&	73	& 9\\
2&Culgoora&	1800--18 MHz&1998--2015&	43 & 2\\
3& e-CALLISTO & 870--45 MHz& 2020--2024 & 280 & 12 \\
 \hline
 \end{tabular}
 \end{table}
\subsection{Instruments}
Spectra corresponding to each Type II burst were obtained from the e-CALLISTO spectrograph. The e-CALLISTO is a network of ground-based radio receiving stations located across the globe. Each radio receiving station uses either a parabolic dish antenna or a log-periodic dipole antenna to continuously monitor the Sun around the globe. Generally, the e-CALLISTO receiver at each station operates in the frequency range of 45--870 MHz, covering the metric and decimetric frequency observations of the Sun originating in the lower and higher corona. Some stations use frequencies below 45 MHz, and others use frequencies above 870 MHz.

The AIA instrument onboard SDO produces full-disk images of the Sun in seven different EUV wavelengths centered at 91 {\AA}, 131 {\AA}, 171 {\AA}, 193 {\AA}, 211 {\AA}, 304 {\AA}, and 335 {\AA}. The EUV images produced are at a pixel resolution of 0.6$''$ with a high time cadence of 12 s {\citep{lemen2012atmospheric}} with the field of view (FOV) of $\sim$1.2$R_{\odot}$. Along with AIA observations, we also used EUVI observations from onboard SECCHI, which observe the Sun in four EUV channels: 171 {\AA}, 195 {\AA}, 284 {\AA}, and 304 {\AA}. These EUV images provide spatial resolution of 1.6$''$ with a minimum cadence of 2.5 minutes. The FOV of the SECCHI/EUVI is $\sim$1.7$R_{\odot}$.

The LASCO/C2 coronagraph onboard SOHO produces white light images every 12 minutes at a spatial resolution of $\sim$23$''$ with a field of view ranging between $\sim$2 to 6R$_\odot$ while LASCO/C3 produces white light images every 12-30 minutes at a spatial resolution of $\sim$112$''$ with a field of view of 3.7 to 30R$_\odot$. We also used SECCHI/COR1 and COR2 onboard STEREO A and B with FOV of is 1.4--4 $R_{\odot}$ and 2.5--15 $R_{\odot}$, respectively. COR1 has a temporal resolution of 5 minutes and a spatial resolution of 3.75$''$, while COR2 observes the Sun at a nominal temporal resolution of 15 minutes with a spatial resolution of 14.7$''$.
\begin{figure*} [!ht]
   \centerline{\hspace*{-0.02\textwidth}
               \includegraphics[width=1.1\textwidth,clip=]{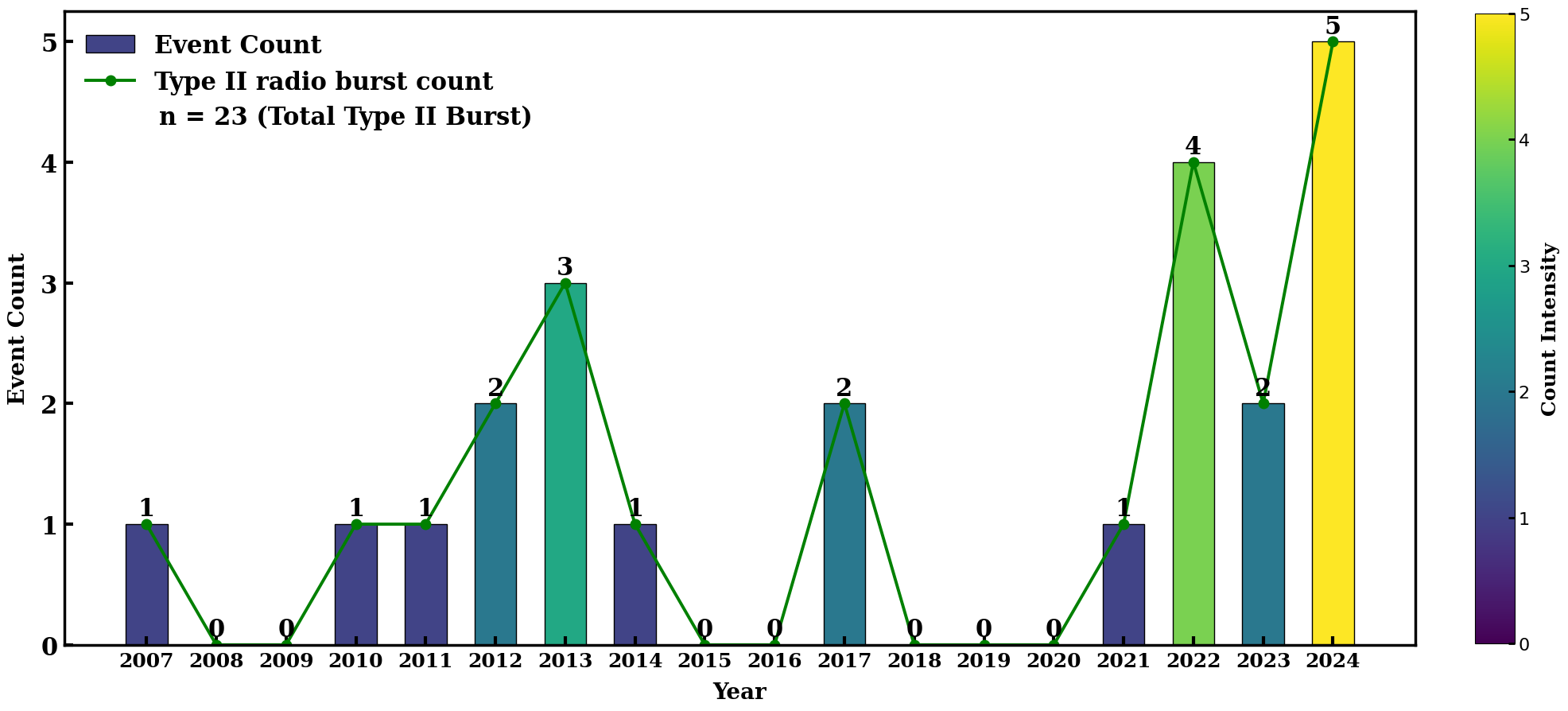}
               \hspace*{-0.03\textwidth}
              }
     \vspace{-0.32\textwidth}   
     \centerline{\Large \bf     
      \hspace{0.0 \textwidth}
      \hspace{0.415\textwidth}
         \hfill}
     \vspace{0.31\textwidth}    

\caption{Histogram of the yearly distribution of Type II solar radio bursts for the period 2007–2024, which are used for the correlation study between CME and Type II radio bursts.}

   \label{Fig1}
   \end{figure*}
%
\section{Measurements of Type II bursts and CMEs}
\label{analysis}
\subsection{Type II radio bursts}
From 396 events of Type II bursts obtained from the three catalogs, we considered only those bursts that could be isolated easily, {\it{i.e.}}, free from any radio frequency interference, and those that didn't overlap with any other bursts. In the present study, we primarily considered the simple and split-band fundamental bursts. Additionally, we considered only the lower-frequency lane of the split-band Type II burst. We identified 50 Type II radio bursts suitable for spectral analysis. The selection mainly aimed at properly estimating the frequency bandwidth of Type II bursts. We further used another selection criterion, as mentioned in Section ~\ref{aw}, and finally selected 23 Type II bursts for our analysis. The details of the selected events as prepared from the Type II burst catalogs are listed in Table \ref{tbl:1}. The yearly distribution of these 23 Type II bursts, covering 2007\,--\,2024 and spanning over solar cycles 24 and 25, is shown in Figure~\ref {Fig1}.
\subsubsection{Estimation of Frequency Bandwidth}
\label{fbw}
%
%

\begin{figure*} [!ht]  
   \centerline{\hspace*{-0.02\textwidth}
               \includegraphics[width=1.\textwidth,clip=]{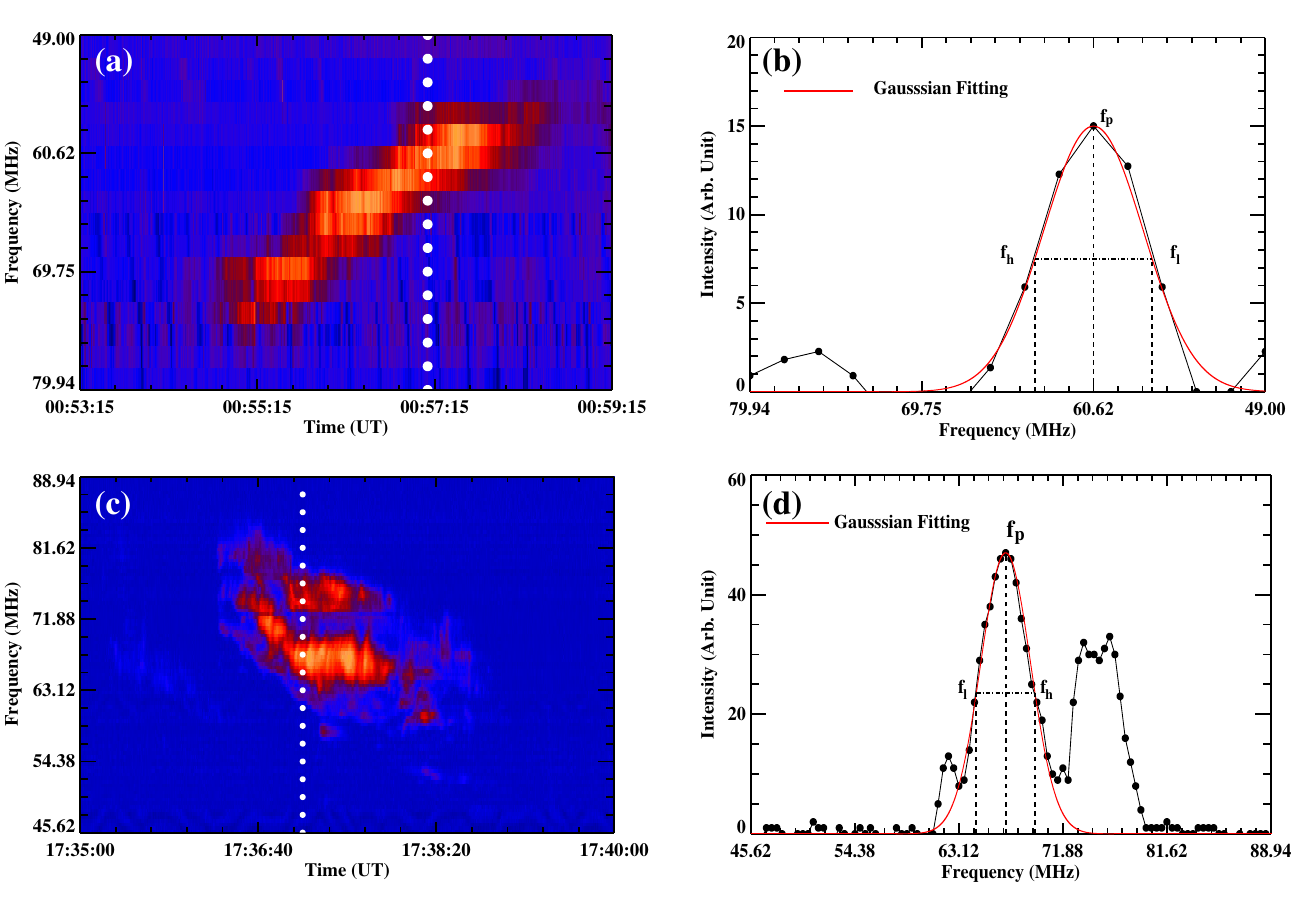}
               \hspace*{-0.03\textwidth}
              }
     \vspace{-0.32\textwidth}   
     \centerline{\Large \bf     
      \hspace{0.0 \textwidth}
      \hspace{0.415\textwidth}
         \hfill}
     \vspace{0.31\textwidth}    

\caption{a) Reduced dynamic spectrum of the Type II burst observed on 31 Dec 2007. The white dots mark the time stamp for the frequency profile of burst intensity as shown in Fig.~\ref{Fig2}b. b) Frequency profile of the burst intensity at the time stamp of 00:57:10 UT. The solid red is the Gaussian fit to the observed burst intensity profile. The long vertical dashed line indicates the central frequency at the profile peak, while the other vertical dashed lines correspond to the two ends of the full-width half maximum of the profile (indicated by a dashed-dotted line). f$_c$ marks the central instantaneous frequency, while f$_h$ and f$_l$ mark the burst's upper and lower frequency limits, respectively. c) Reduced dynamic spectrum of the Type II burst observed on 14 May 2024. The white dots mark the time stamp for the frequency profile of burst intensity as shown in Fig.~\ref{Fig2}d. d) Frequency profile of the burst intensity of the split-band Type II burst at time stamp 17:37:05 UT. The rest is the same as Fig.~\ref{Fig2}b.}

   \label{Fig2}
   \end{figure*}
%
Figure \ref{Fig2}a shows a reduced dynamic spectrum of a Type II burst on 31 Dec 2007 observed by the e-Callisto network of stations. We prepared the reduced dynamic spectrum by subtracting the median value of the radio flux obtained for each frequency row from the original radio flux of each column {\citep{singh2019automated}}.
As shown in Fig.~\ref{Fig2}a, the burst features are now clearly isolated. The burst identified in Fig.~\ref{Fig2}a depicts the typical characteristics of a Type II burst, which usually drifts slowly from a higher to a lower frequency regime. The isolated burst in Fig.~\ref{Fig2}a is a single-lane fundamental Type II burst. Figure~\ref{Fig2}c shows a reduced dynamic spectrum of a split-band fundamental Type II burst on 14 May 2024.

We used a method, as illustrated in Fig.~\ref{Fig2}b and Fig.~\ref{Fig2}d, to estimate the frequency bandwidth of the Type II burst. The technique is based on a method used by {\cite{aguilar2005universal}}.
In this technique, we obtained the frequency profile of the observed radio burst flux at different time stamps after reducing the dynamic spectrum for each selected burst. Figure \ref{Fig2}b shows such frequency profiles for a simple Type II burst. The solid black curve represents the observed frequency profile of the burst intensity at a given timestamp. Then, we performed a Gaussian fit to the observed flux profile as a function of frequency, and the best-fit Gaussian was overplotted in red. The full width at half maximum (FWHM) of the Gaussian fit represents the instantaneous frequency bandwidth of the concerned Type II burst at that time stamp.
Similarly,  flux profiles were obtained as a function of frequency for different time stamps of the burst, and a Gaussian fit was performed to estimate the instantaneous frequency bandwidth. Finally, the frequency bandwidth of the Type II burst was determined as the average of all instantaneous bandwidths. In addition, using the Gaussian fit to the observed frequency profiles of the burst intensity, the peak frequency, and the lower and higher frequency limits of the burst at the corresponding time stamp were obtained, which are indicated by dotted vertical lines in Fig.~\ref{Fig2}b and Fig.~\ref{Fig2}d. We also calculated the relative frequency bandwidth (RBW) of each Type II burst. The RBW is obtained by dividing the instantaneous bandwidth by the central or peak frequency of the corresponding burst.

Table \ref{tbl:2} lists all 23 Type II burst events, providing details of the event date, frequency interval, frequency bandwidth, and RBW. The frequency range, bandwidth, and RBW of the finally selected 23 Type II burst events are listed in Columns 4 and 5 of Table~\ref{tbl:2}. It is evident from Table~\ref{tbl:2} that the values of RBW obtained in this study for coronal Type II bursts are in the range 0.03\,--\,0.24. The values are consistent with the values of RBW reported by {\cite{ramesh2022new}} for coronal Type II bursts. The values obtained here for coronal Type II bursts agree with the average value of 0.23$\pm$0.13 reported by {\cite{aguilar2005universal}} for interplanetary Type II bursts. 
\subsubsection{Estimation of Type II burst heights}
\label{Burst_height}
Burst heights can be estimated accurately from radio imaging observations of Type II emission sources, accounting for scattering and refraction. When radio waves travel through free space from the source region to the observer, the observed and actual positions and sizes of the source don't differ. However, when radio waves (here, Type II emissions) travel through the solar corona and interplanetary space, the observed positions and sizes of the sources differ from their true values. This is due to the scattering and refraction of radio waves caused by variations in the density of the intervening medium, a phenomenon known as {radio wave propagation effects}. Since no radio images were available for our Type II burst events, we used the radio dynamic spectrum and coronal density model to estimate the height of Type II burst emission. The radio dynamic spectrum provides only the intensity information of the burst at a given frequency and time. Thus, it isn't easy to account for the effects of scattering and refraction on Type II sources. Hence, we kept the estimation of the height of Type II sources simple. We used the start frequency of the burst at the onset time of the Type II burst event from the e-Callisto radio dynamic spectra. The start frequency used here corresponds to the burst's lower frequency, determined by a Gaussian fit to the burst-intensity frequency profile at the burst's onset. This Gaussian fit technique has been described in Section~\ref{fbw} and depicted in Fig.~\ref{Fig2}b. Subsequently, using the fundamental frequency relation to plasma density, $f{_{p}}\sim8.98\sqrt{n_{e}}$,  we converted the obtained start frequency at the onset time of the burst into the corresponding plasma density. Next, to get the corresponding height, we used three density models: Newkirk density model {\citep{newkirk1961solar}}, Saito density model {\citep{saito1970non}}, and Leblanc density model {\citep{leblanc1998tracing}} represented, respectively, by equations (1), (2), and (3).

\begin{equation}
n_{e} = 4.2\times10^{4}\times10^{4.32R_\odot/r}
\end{equation}

\begin{equation}
n_{e} = 3.1\times10^{8}r^{-16} + 1.58\times10^{8}r^{-6} + 0.025\times10^{8}r^{-2.5}
\end{equation}

\begin{equation}
n_{e} = 3.3\times10^{5}r^{-2} + 4.1\times10^{6}r^{-4} + 8.0\times10^{7}r^{-6}
\end{equation}

We estimated an average burst height using the heights obtained from the three coronal density models. {The burst heights corresponding to each Type II burst obtained using the three coronal density models and the average burst height are listed in Table~\ref{tbl:3}.}\\

\begin{figure} [!h]

    \centerline{\hspace*{0.04\textwidth}
                \includegraphics[width=1.5\textwidth,clip=]{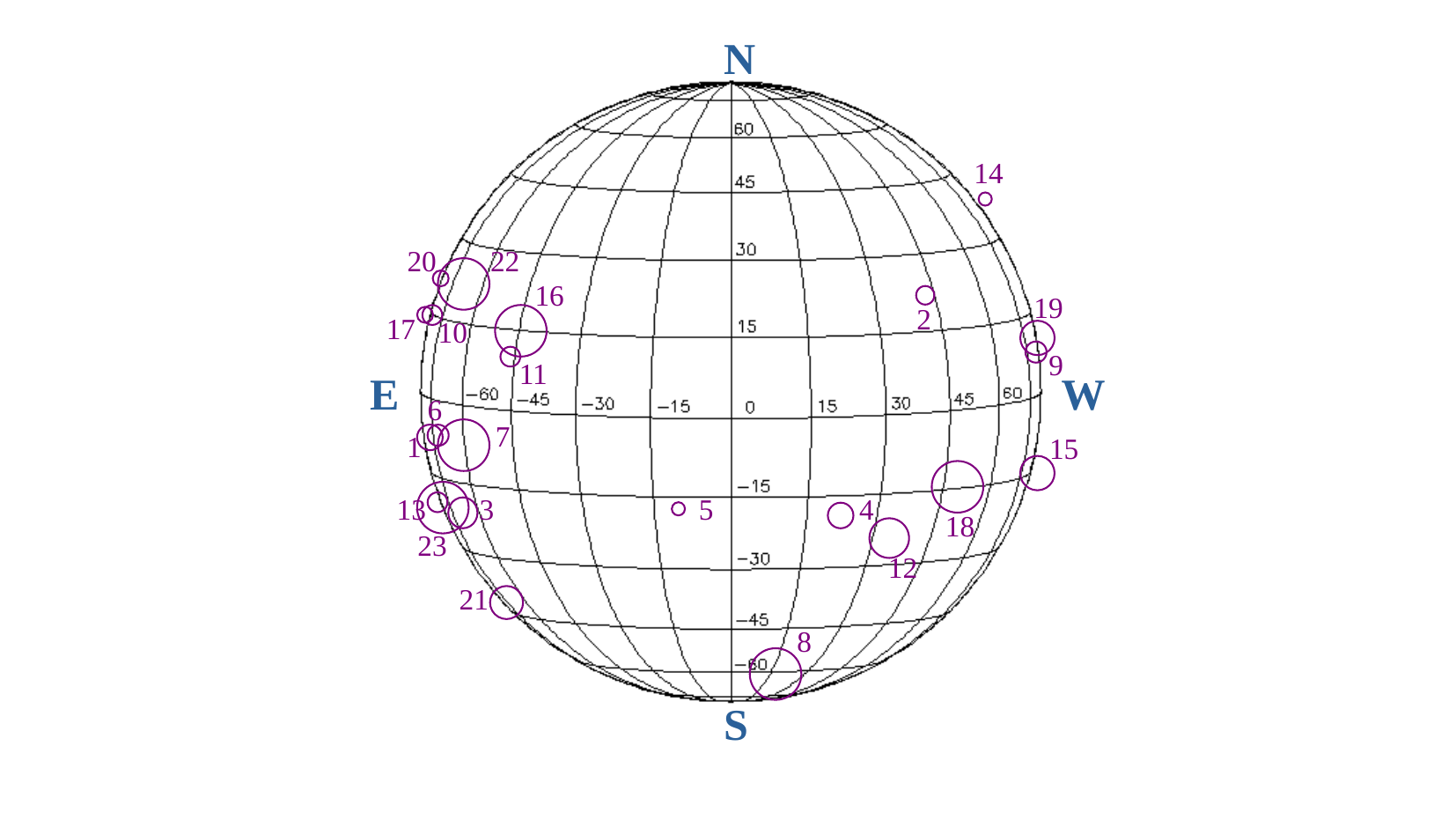}
                \hspace*{0.03\textwidth}
               }
    \vspace{-0.53\textwidth}
      \centerline{\Large \bf
    \hspace{0.0 \textwidth}
       \hspace{0.415\textwidth}
         \hfill}
      \vspace{0.5\textwidth}
\caption{Locations of source ARs of the CMEs associated with the Type II burst events spread across different longitudes and latitudes over the Sun. The circle with the number indicates the CME associated with each Type II burst event, provided in Table~\ref{tbl:2}. The different sizes of the circle represent the different angular widths of CMEs, with the bigger circle indicating the higher angular width and so on.}

   \label{Fig3}
   \end{figure}

%
\begin{landscape}
\begin{table}
 \caption{List of all 23 Type II burst events used in this study, providing the details of the events, the properties of Type II bursts, and the properties of the associated CMEs. \\}

 {Col.(1): Serial number, Col.(2): Event date, Col.(3): Start time of the burst,
 Col.(4): Frequency range of the burst, \\
 Col.(5): Average Frequency bandwidth/Relative Frequency bandwidth, Col.(6): Location of the active region,\\
 Col.(7): Apparent angular width of CME, and Col.(8): Instantaneous angular width of CME.}

\label{tbl:2}



\begin{tabular}{cccccccc}     
\hline

 &  & & & & &CME Apparent  & CME Ins.\\
Serial& Date & Burst & $f_s - f_e$ & $\Delta{f}$/RBW&AR &angular width & angular width\\
No. & DD/MM/YYYY & Time(UT) & (MHz) & (MHz) & Location&(Deg.)&(Deg.)\\
\hline

1&31/12/2007(SL)&00:55&80-50&6.19/0.10&S08E81&164&21.86\\
2&12/06/2010(SL)&00:58&80-60&7.69/0.12&N23W43&119&33.00\\
3&08/03/2011(SL)&03:45&85-50&7.012/0.10&S21E72&260&54.64\\
4&08/06/2012(SL)&03:06&125-110&11.79/0.11&S19W21&119&20.36\\
5&02/07/2012(SL)&05:10&115-55&12.85/0.20&S17E10&90&23.32\\
6&25/10/2013(SL)&03:03&85-45&16.64/0.28&S07E78&121&31.34\\
7&25/10/2013(SL)&08:01&85-35&10.09/0.25&S06E73&360&25.41\\
8&19/11/2013(SB)&10:27&85-45&10.82/0.19&S70W14&360&30.20\\
9&20/04/2014(SL)&08:19&85-50&7.99/0.12&N10W87&123&33.58\\
10&18/04/2017(SL)&09:43&80-50&13.69/0.24&N15E85&101&31.97\\
11&12/09/2017(SL)&07:32&60-20&7.58/0.31&N08E48&96&20.83\\
12&28/09/2021(SB)&06:23&30-20&2.37/0.03&S26W36&265&41.90\\
13&12/02/2022(SB)&08:34&150-110&12.60/0.10&S19W84&102&29.73\\
14&18/02/2022(SL)&19:17&60-40&10.15/0.22&N40W90&58&36.27\\
15&14/03/2022(SL)&17:21&85-30&9.23/0.22&S15W90&225&29.63\\
16&13/06/2022(SB)&03:25&35-25&2.71/0.08&N15E45&360&61.61\\
17&14/01/2023(SL)&12:47&80-65&3.4/0.05&N15E90&71&44.47\\
18&10/07/2023(SB)&03:40&30-15&2.8/0.13&S15W48&360&36.54\\
19&14/01/2024(SL)&11:46&55-30&5.02/0.11&N01W89 &215&49.81\\
20&29/01/2024(SL)&10:23&80-60&9.95/0.16&N19E87&71&21.74\\
21&21/02/2024(SL)&02:16&55-35&4.34/0.09&S40E84&188&50.05\\
22&14/05/2024(SB)&17:36&80-60&4.76/0.07&N19E72 &360&42.04\\
23&27/05/2024(SL)&07:12&35-20&5.5/0.24&S18E89&360&45.25\\

 \hline
\end{tabular}
\\

SL = Single lane Type II burst\\
SB = Split band Type II burst\\
$f_{s}$ = Start frequency of Type II burst\\
$f_{e}$ = End frequency of Type II burst\\
$\Delta{f}$ = Type II burst frequency bandwidth\\
RBW = Relative bandwidth\\
AR = Active region\\

\end{table}
\end{landscape}
%

\subsection{CMEs associated with Type II radio bursts}
\subsubsection{Estimation of CME angular width}
\label{aw}
The solar surface distribution of the source ARs of the CMEs associated with the Type II bursts is shown in Fig.~\ref{Fig3}. The circles with numbers indicate the CME locations associated with each Type II burst as listed in Table~\ref{tbl:2}, and the size of the circle indicates its angular width. The angular widths used were from the SOHO/LASCO CME catalog and are also listed in Table~\ref{tbl:2}. In Fig.\ref{Fig3}, the bigger the circle, the higher the angular width of the CME. From Table~\ref{tbl:2} and Fig~\ref{Fig3}, it is seen that the source locations of the CMEs associated with the Type II bursts are in the latitude range of -30$^{\circ}$ to +30$^{\circ}$. On the other hand, the source locations span the entire longitude range from 0$^{\circ}$ to 360$^{\circ}$. Using the values of CME apparent angular width and the average frequency bandwidth from Table~\ref{tbl:2}, an anti-correlation with correlation coefficients (cc) of -0.46 is observed when we made a plot 
between the CME apparent angular widths and the average frequency bandwidths {as shown in Fig.~\ref{Fig4}}. The correlation obtained using the apparent angular widths of the associated CMEs, which are distributed over a wide range of heliographic longitudes as shown in Fig.\ref{Fig3}, may produce misleading results. Hence, in this study, we estimated the 3D angular width of CMEs associated with the shortlisted Type II burst using a forward-fitting GCS model.
\begin{figure} [!h]

\centerline{\hspace*{0.02\textwidth}

               \includegraphics[width=\textwidth,clip=]{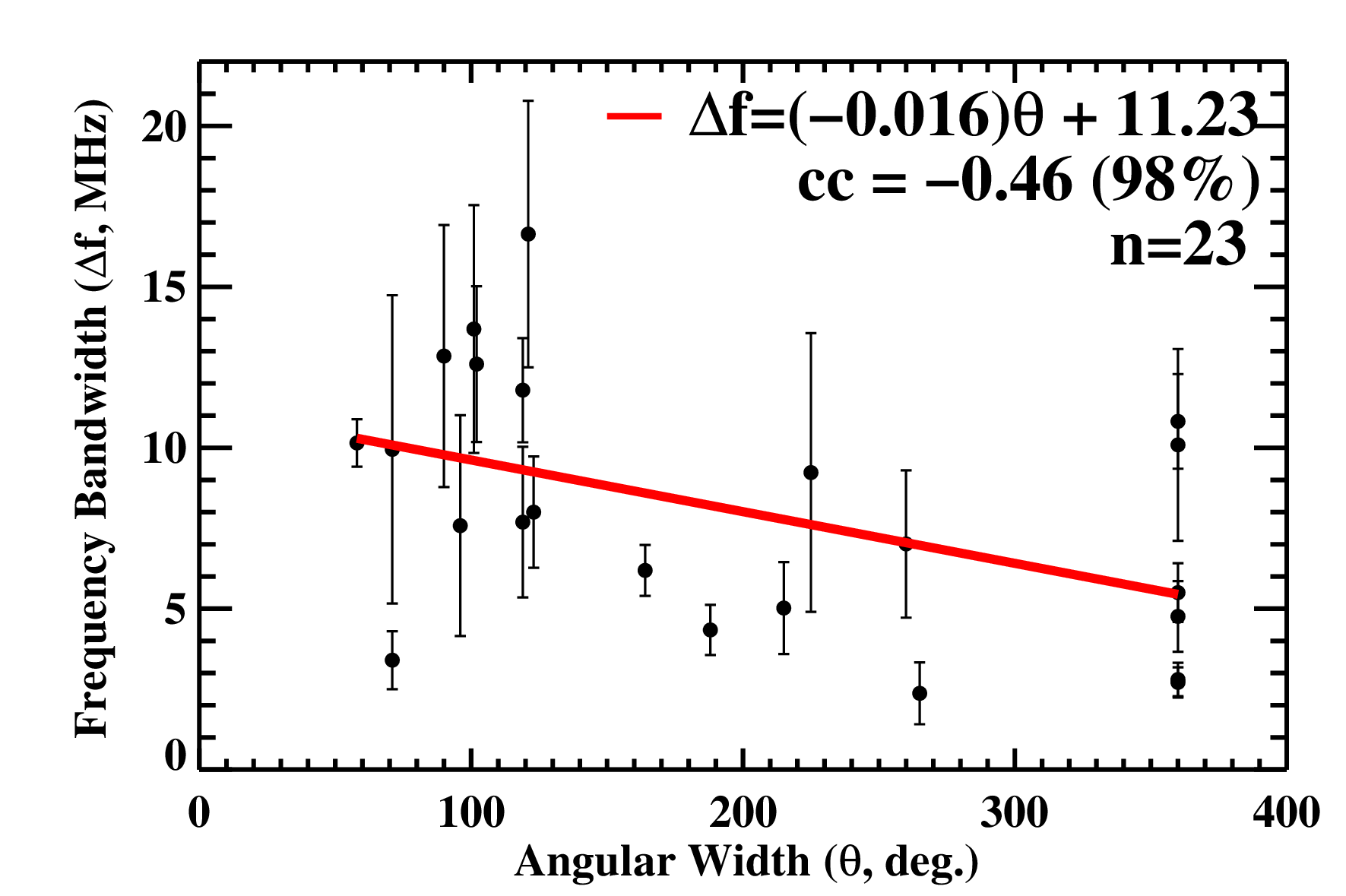}
               \hspace*{0.04\textwidth}%

            }%

     \vspace{-0.325\textwidth}   
     \centerline{\Large \bf     
      \hspace{0.08 \textwidth}
      \hspace{0.765\textwidth}
        \hfill}
     \vspace{0.29\textwidth}    

\caption{The frequency bandwidth of the 23 Type II bursts observed with e-Callisto spectrograph during 2007\,--\,2024 in the frequency range 150\,--\,20 MHz as a function of SOHO/LASCO angular width of the associated CMEs. The functional forms of the linear fit and correlation coefficients (cc) with confidence levels are indicated in the top right-hand corner.}

   \label{Fig4}
   \end{figure}


In the GCS model, the estimation of the CME angular width can be made by reconstructing a 3-D shape of a CME, provided a CME has been observed at two vantage points. When multi-viewpoint observations of CMEs corresponding to our shortlisted Type II bursts were not available, we considered those Type II bursts for which the associated CMEs erupted close to the limb as observed by either LASCO or STEREO. The CMEs detected by LASCO or STEREO are observed through Thomson scattering due to the free electrons present in the solar corona. In the case of limb CMEs, they are directly perpendicular to the LOS. Thus, the efficiency of Thomson scattering is maximized, and the three-part structure of the CMEs could be clearly detected. The forward-fitting GCS model can then be successfully employed to reconstruct the 3D shapes of CMEs and determine their angular widths. From our shortlisted 50 Type II bursts, we found 23 Type II burst events for which the angular width of the associated CMEs could be estimated using the GCS model.


\begin{figure} [!h]

    \centerline{\hspace*{0.04\textwidth}
                \includegraphics[width=\textwidth,clip=]{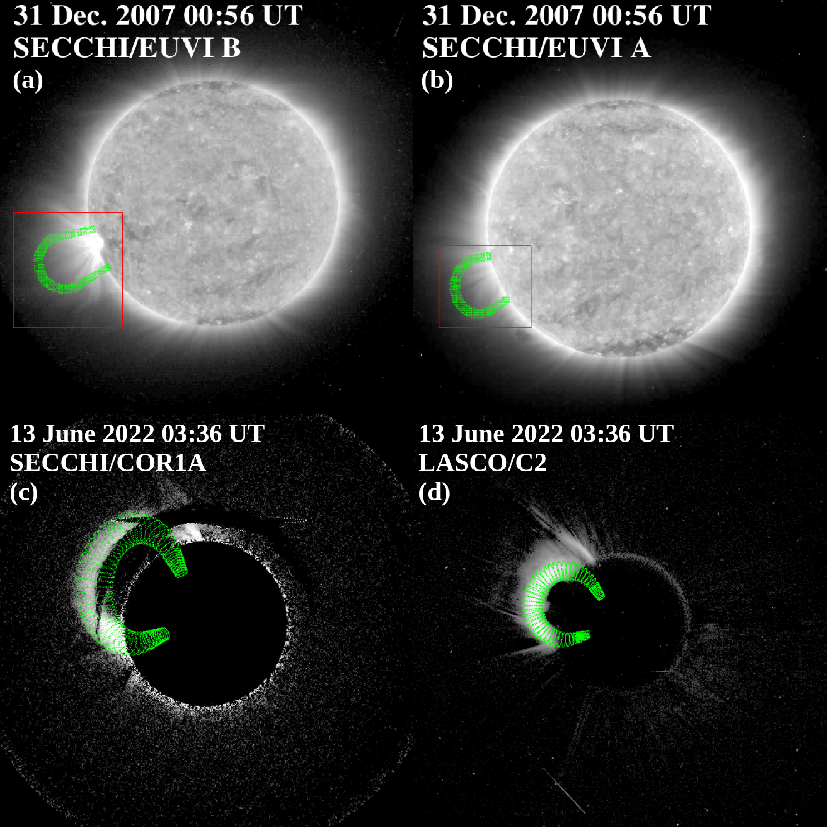}
                \hspace*{0.03\textwidth}
               }
    \vspace{-0.53\textwidth}
      \centerline{\Large \bf
    \hspace{0.0 \textwidth}
       \hspace{0.415\textwidth}
         \hfill}
      \vspace{0.5\textwidth}

\caption{a) The GCS fit to the filtered image of the CME as observed in STEREO/EUVI B at 00:56 UT on 31 Dec. 2007. b) The GCS fit to the filtered image of the CME as observed in STEREO/EUVI A at 00:56 UT on 31 Dec. 2007. c) The GCS fit (green wireframe) to the running difference image of the CME from SECCHI/COR1A at 03:36 UT on 13 June 2022. d) The GCS fit to the running difference image of CME from LASCO/C2 at 03:36 UT on 13 June 2022.
}

   \label{Fig5}
   \end{figure}

Using the GCS model, we fitted a sequence of difference images that depict the expansion of CMEs at different timestamps. The fitting provides parameters such as, longitude ($\phi$), latitude ($\theta$), tilt angle($\gamma$), height (H in $R_{\odot}$), aspect ratio ($\kappa$), half angle ($\alpha$) at different time stamps. Angular width was then calculated as $\theta = 2(\alpha + \delta)$ for each time stamp, where $\delta = arcsin(\kappa)$. We estimated the angular width of the CMEs using the data from STEREO COR1, COR2 (A, B), and/or LASCO (C2, C3). Generally, the metric Type II bursts appear during the increase of the CME angular widths. Hence, we estimated the angular width at the time of the burst, which we refer to as the instantaneous angular width. 

\begin{figure} [!ht]

    \centerline{\hspace*{0.01\textwidth}
                \includegraphics[width=1.05\textwidth,clip=]{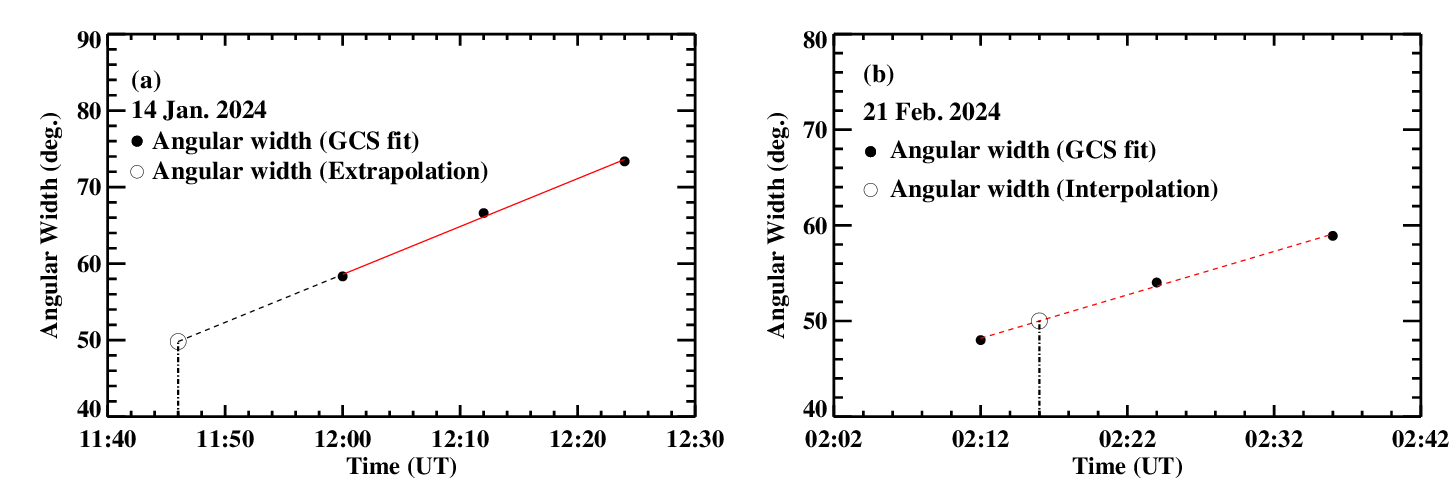}
                \hspace*{0.03\textwidth}
               }
    \vspace{-0.53\textwidth}
      \centerline{\Large \bf
    \hspace{0.0 \textwidth}
       \hspace{0.415\textwidth}
         \hfill}
      \vspace{0.49\textwidth}

\caption{a) Estimation of the instantaneous CME angular width using an extrapolation method for a burst event on 14 Jan. 2024. b) Estimation of the instantaneous CME angular width using an interpolation method for a burst event on 21 Feb. 2024. The solid black circles represent the estimated angular width of the CME, determined from GCS fits to coronal images at different timestamps. The red straight line is a linear fit to the instantaneous CME angular widths obtained. The dashed line in panel a is the extrapolation of the linear fit up to the burst onset time, marked by a vertical dashed line. An open circle marks the estimated instantaneous CME angular width at the burst onset time.}
    \label{Fig6}
    \end{figure}
%
\subsubsection{Instantaneous Angular Width}
\label{iaw}
We employed the GCS fit to estimate the instantaneous angular width of the CME at the burst onset time. Figure~\ref{Fig5}(a, b) shows the GCS fit for a limb event on 31 Dec. 2007. The panels (a, b) of Fig.~\ref{Fig5} show the images from SECCHI/EUVI B and A, respectively, depicting the initiation of the CME eruptions close to the burst onset time at 00:55 UT, along with the GCS fit to the erupted CMEs. Using the GCS fit parameters, we then estimated the instantaneous angular width at the burst onset time for the 31 Dec. 2007 event. As in the case of 31 Dec. 2007, for the other six events with white-light images available at the burst onset, we estimated the instantaneous angular width using the GCS fit to the images at that time.

In the remaining event cases, when there were no imaging observations at the burst onset time, we used a sequence of white-light coronal images of CMEs from SECCHI/EUVI, SECCHI/COR1, and LASCO/C2 at different timestamps close to the burst onset time and employed the GCS fit to estimate the angular width at those timestamps. We thereafter used an extrapolation/interpolation technique on the fitted data points to obtain the instantaneous angular width. Figure~\ref{Fig5}(c, d) shows one such event on 13 Jun. 2022, for which imaging observations were not available at the burst onset time. Hence, we fitted CMEs using the GCS model at different timestamps to obtain their angular widths. Here, we have shown the GCS fit at one of the closest available times, 03:36 UT, for this event. For the sixteen burst events, we used the back-extrapolation method in nine cases. In contrast, in seven event cases, we used the interpolation method to obtain the CME angular width at the burst onset time.

Figure ~\ref{Fig6}{a} shows a plot of the angular widths of an event on 14 Jan. 2024 at the different time stamps, marked by the filled black solid circles. For this event case, we obtained the instantaneous angular widths by back-extrapolating the fitted data points, which were estimated using the GCS fit to the coronal images. The red straight solid line in Fig.~\ref{Fig6}{a} is a linear fit to the estimated angular widths of the CMEs at different time stamps other than the burst onset time. To calculate the angular width at the burst onset time, indicated by a vertical line, we back-extrapolated the linear fit to the burst onset time, as shown by the dashed line in Fig.~\ref{Fig6}{a}. The estimated instantaneous angular width using the extrapolation of the linear fit is indicated by an open circle in Fig.~\ref{Fig6}{a}. Similarly, Fig.~\ref{Fig6}b shows a plot of the angular widths of an event on 21 Feb. 2024 at the different time stamps, where we used an interpolation to the fitted data points to obtain the angular width at the burst onset time. The black open circle represents the obtained CME angular width at the burst onset time, which is the instantaneous angular width of the CME for that event.

The estimated instantaneous angular widths of CMEs associated with the Type II bursts are listed in Column 8 of Table~\ref{tbl:2}.
\subsubsection{Estimation of CME Heights}
\label{CME_height}

To estimate the CME heights, from the finally shortlisted 23 Type II burst events, images were obtained for 8 events from either SDO/AIA or LASCO/C2, depending on the images available at the onset time of the Type II bursts. On the other hand, for the remaining 15 events, images from SDO/AIA or LASCO/C2 were not available at the time of the burst; therefore, we used images from either SECCHI/EUVI or SECCHI/COR1 on board STEREO A and B. For illustrative purposes, we discussed the case studies of the 12 Jun. 2010 burst.

\begin{figure*} [!ht]

  \centerline{\hspace*{-0.06\textwidth}
            \includegraphics[width=1.05\textwidth,clip=]{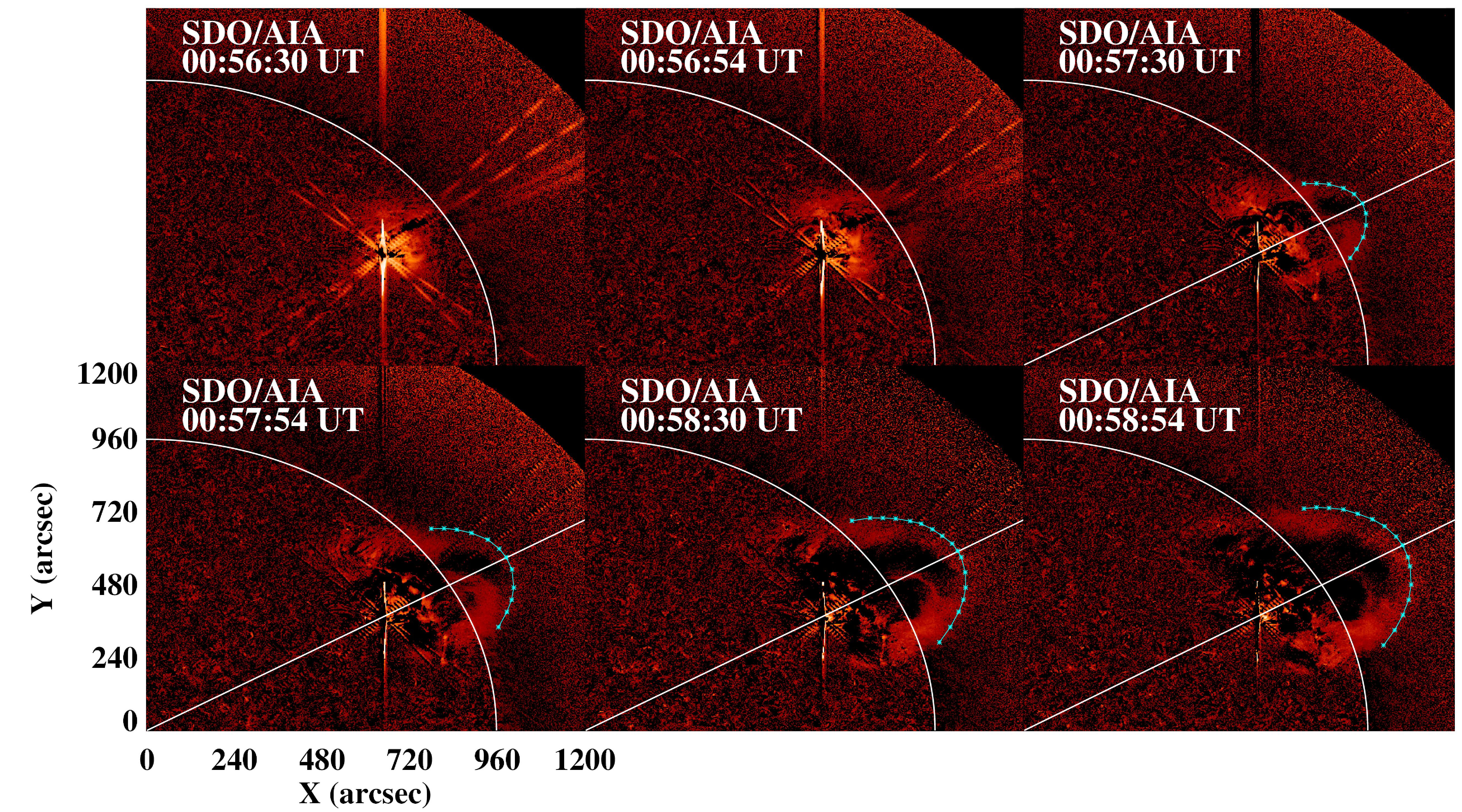}
                \hspace*{-0.03\textwidth}
             }
      \vspace{-0.33\textwidth}
      \centerline{\Large \bf

       \hspace{0.0 \textwidth}
       \hspace{0.415\textwidth}
          \hfill}
      \vspace{0.31\textwidth}

\caption{A sequence of running difference images of the Sun from SDO/AIA, showing the eruption and evolution of CME associated with the Type II burst event of 12 Jun. 2010. The timestamps for each image are displayed in the top-left corner of each panel. The thick white circle in each panel represents the solar limb, whereas the cyan asterisks shown in the last four panels are the ellipsoid fit to the CME fronts. A solid line in white in each of the last four panels is a slit used from the Sun's center to the apex of the CME front to estimate the corresponding CME height at each time stamp. }

    \label{Fig7}
    \end{figure*}
%
\begin{figure*} [!ht]   

  \centerline{\hspace*{-0.03\textwidth}
            \includegraphics[width=1.1\textwidth,clip=]{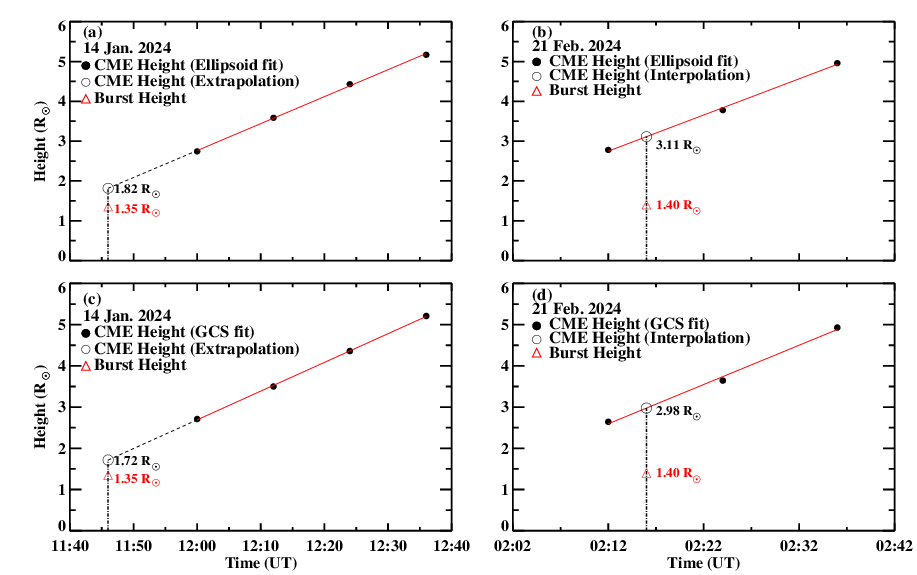}
                \hspace*{-0.03\textwidth}
             }
      \vspace{-0.33\textwidth}
      \centerline{\Large \bf

       \hspace{0.0 \textwidth}
       \hspace{0.415\textwidth}
          \hfill}
      \vspace{0.31\textwidth}

\caption{a, c) Estimation of the CME height using the extrapolation method for the burst event on 14 Jan. 2024. b, d) Estimation of the CME height using the interpolation method for the burst event on 21 Feb. 2024. The CME height is measured from the center of the Sun using the ellipsoid (panel a, b) and GCS fits (panel c, d), respectively. The black, solid-filled circles represent the estimated CME heights from the ellipsoid/GCS fits, which fit the coronal images of the CMEs at different time stamps. The straight red line is a linear fit to the CME heights obtained. The dashed line is the extrapolation of the linear fit up to the burst onset time, marked by a vertical dashed line. The estimated CME height, obtained by extrapolating and interpolating the data points, is marked by an open circle. In contrast, the burst height obtained for the same event is marked by a red triangle for comparison.}

    \label{Fig8}
    \end{figure*}

Figure~\ref{Fig7} shows a sequence of running difference images from SDO/AIA, where we can see a CME eruption and its successive evolution traveling as a disturbance from the Sun. The CME is associated with the Type II burst event observed at 00:58 UT on 12 Jun. 2010. The eruption of CME occurring in the north-west limb of the Sun is clearly seen in the running difference image at 00:57:30 UT, and its successive evolution is further evident from the series of snapshots between 00:56:30 UT\,--\,00:58:54 UT in Fig.~\ref{Fig7}.
The thick white circle in each panel represents the solar limb. To estimate the CME height, an ellipsoid was fitted to manually selected data points at the CME front in images where the front is discernible. For example, for the event of Jun. 12, 2010, as shown in Fig.~\ref{Fig7}, we estimated the height of the CME for four time instants corresponding to the running difference images marked with cyan asterisks in Fig.~\ref{Fig7}, which show the ellipsoid fits. Next, a slit was used from the center of the Sun to the apex of the CME front, as shown in Fig.~\ref{Fig7}. The height of the CME is here referred to as the distance of the slit from the center of the Sun to the apex point of the CME. Similarly, the CME heights for the other burst events were estimated and are listed as CME heights estimated using ellipsoid fits in Column 7 of Table~\ref{tbl:3}. In cases with 9 events, we directly used white-light/EUV images at the burst onset to estimate the CME height at that time. However, in the remaining 14 events, we used either interpolation or extrapolation to estimate the CME height at the time of the burst. Of these 14 events, in 5 cases the CME height was determined by extrapolation, as shown in Fig.~\ref{Fig8}a for a burst event on 14 Jan. 2024. The black-filled circle marks the CME heights obtained using the ellipsoid fit to the coronal images of the CMEs available close to the burst period. The red solid line is a linear fit to these observed data points. The dashed line represents the back extrapolation up to the burst onset time, and the black open circle indicates the CME height obtained at the time of the burst. In cases of the other nine events, the CME height was obtained using an interpolation method as shown in Fig.~\ref{Fig8}b for a burst event of 21 Feb. 2024.

In addition to estimating the CME height from direct images, as explained earlier, it was also obtained from GCS fits, which provide CME height information at the time of the burst. As mentioned, when coronal images of the CMEs were unavailable at the time of the burst, we used either extrapolation or interpolation to estimate the CME heights at the onset time of the burst. Figures ~\ref{Fig8}c and ~\ref{Fig8}d show, respectively, the height estimation of the CMEs using the backward extrapolation and interpolation methods for burst events of 14 Jan. and 21 Feb., 2024. In such cases, the CME heights were estimated using timestamps other than the burst onset time when coronal images of the CMEs were available, as indicated by the black-filled circles in Figs. ~\ref{Fig8}c and ~\ref{Fig8}d. The red line is a linear fit to these observed data points. To estimate the CME height at the burst onset time (indicated by an open circle), we back-extrapolated or interpolated the data points to that time. The CME heights obtained, thus, using the GCS fits are listed as CME heights estimated using GCS in Column 3 of Table~\ref{tbl:3}.
\section{Results}
\label{res}
\subsection{Relation Between Frequency Bandwidth and Angular Width}
\label{bwaw}
We plotted the angular widths obtained from the GCS fits against the average frequency bandwidth of 23 Type II bursts. Figure~\ref{Fig9} shows the correlation between the frequency bandwidth and the instantaneous angular width. The correlation coefficient is cc=-0.62. The filled circles in Fig.~\ref{Fig9} are observed data points with 1 $\sigma$ error bars. Further, to show the dependence of frequency bandwidth (${\Delta}f$) on the angular width ($\theta$), the observed data points were fitted with linear least squares fits, as shown by solid red lines. It is further evident from Fig.~\ref{Fig9} that the angular width of the CMEs and the frequency bandwidth of the Type II bursts are anti-correlated.

\begin{figure} [!h]

\centerline{\hspace*{0.02\textwidth}

               \includegraphics[width=\textwidth,clip=]{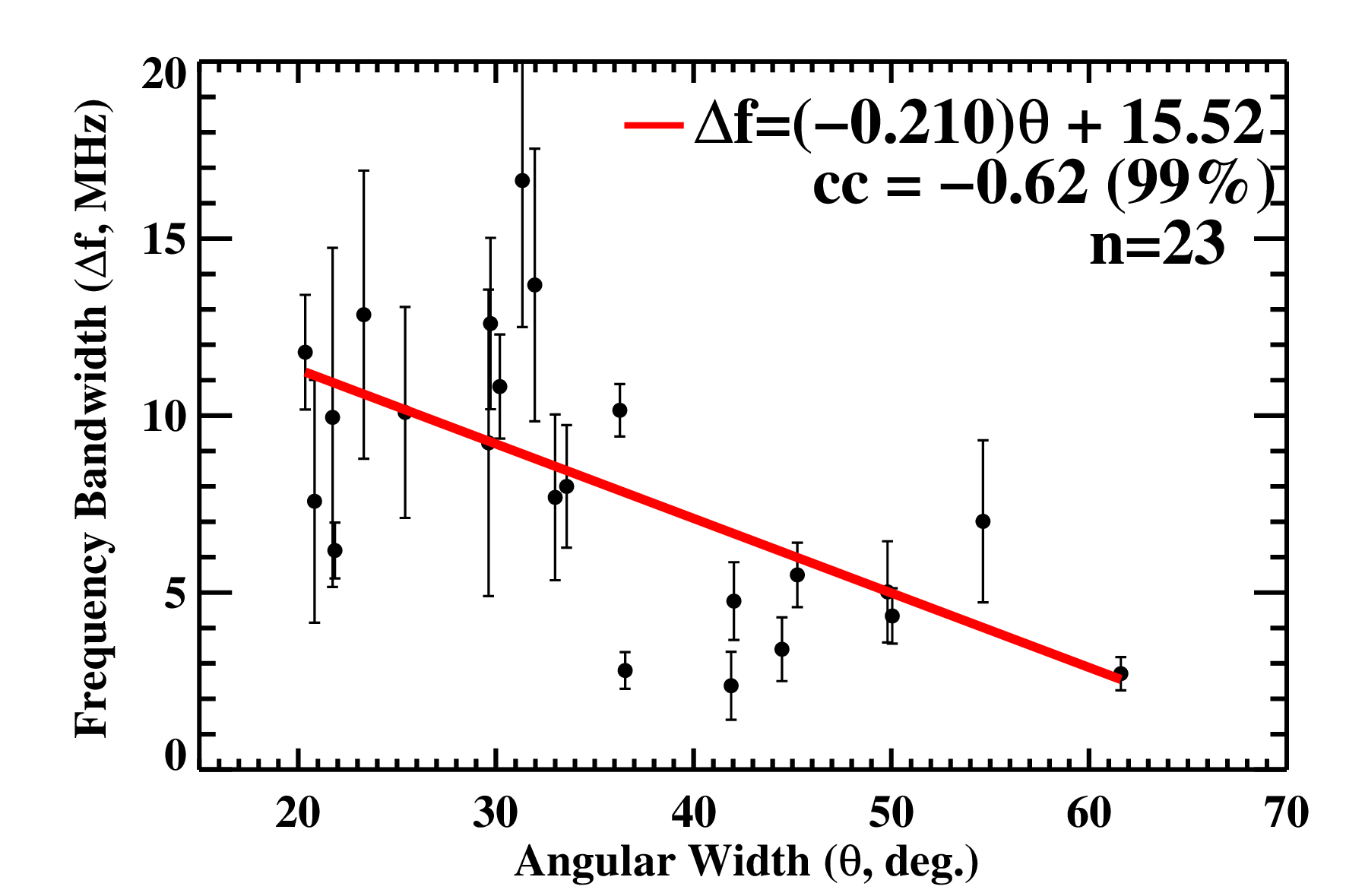}
               \hspace*{0.04\textwidth}%

            }%

     \vspace{-0.325\textwidth}   
     \centerline{\Large \bf     
      \hspace{0.08 \textwidth}
      \hspace{0.765\textwidth}
        \hfill}
     \vspace{0.29\textwidth}    

\caption{The frequency bandwidth of the 23 Type II bursts observed with e-Callisto spectrograph during 2007\,--\,2024 in the frequency range 150\,--\,20 MHz as a function of instantaneous GCS angular width of the associated CMEs. The functional forms of the linear fit and correlation coefficients (cc) with confidence levels are indicated in the top right-hand corner.}

   \label{Fig9}
   \end{figure}


\subsection{Comparing burst heights and corresponding CME heights}
\label{h_compare}
Since we aim to find out whether the height of the burst at the onset of the burst is greater or less than the CME height, we plotted the CME heights obtained at different time stamps and compared them with the average burst height at the onset time of the burst for each Type II burst event. Figure ~\ref{Fig8} shows the height-time plot for the burst events of 14 Jan. and 21 Feb., 2024. The black-filled circle marks the CME heights obtained using the ellipsoid/GCS fit to the available CME images. 
%

\begin{landscape}
\begin{table}
 \caption{Details of the estimated heights of all 23 Type II burst events and their associated CMEs.\\}

{ Col.(1): Serial number, Col.(2): Event date, Col.(3): Height of the CME, Col.(4): Height of the Type II burst using Newkirk Model (NM),\\
Saito Model (SM), Leblanc Model (LM), and the average burst height, Col.(5): The difference between the height of the CME and the\\
Type II burst.}

\label{tbl:3}

\begin{tabular}{cccccccc}     
\hline

Sr.  &Date &CME Height($r_{CME}(R_{\odot}))$&& Burst height $r_{TypeII}(R_{\odot})$&& & $\Delta{r}$\\
No.&DD/MM/YYYY&(GCS/Ellipsoid Fits)&NM&SM&LM&Avg. Height&$r_{CME} - r_{TypeII}(R_\odot) $\\

\hline

 1&31/12/2007(SL)&1.57/1.50&1.37&1.22&1.06&1.22&0.35/0.28\\
 2&12/06/2010(SL)&1.63/1.19&1.36&1.22&1.05&1.21&0.42/-0.02\\
 3&08/03/2011(SL)&1.51/1.61&1.34&1.19&1.03&1.19&0.32/0.42\\
 4&08/06/2012(SL)&1.36/1.34&1.21&1.10&0.91&1.07&0.29/0.27\\
 5&02/07/2012(SL)&1.43/1.40&1.38&1.23&1.07&1.23&0.20/0.17\\
6&25/10/2013(SL)&1.78/1.51&1.36&1.22&1.06&1.19&0.59/0.32\\
7&25/10/2013(SL)&1.44/1.30&1.48&1.32&1.16&1.32&0.12/-0.02\\
8&19/11/2013(SB)&1.68/1.72&1.34&1.20&1.02&1.19&0.49/0.53\\
9&20/04/2014(SL)&1.68/1.62&1.35&1.21&1.04&1.20&0.48/0.42\\
10&18/04/2017(SL)&1.77/1.34&1.48&1.32&1.16&1.32&0.45/0.02\\
11&12/09/2017(SB)&1.78/1.59&1.60&1.43&1.27&1.43&0.35/0.16\\
12&28/09/2021(SB)&1.89/2.06&1.93&1.70&1.52&1.72&0.17/0.34\\
13&12/02/2022(SB)&1.29/1.30&1.14&1.05&0.83&1.01&0.28/0.29\\
14&18/02/2022(SL)&2.00/1.48&1.48&1.32&1.16&1.32&0.68/0.16\\
15&14/03/2022(SL)&1.33/1.15&1.33&1.20&1.02&1.18&0.15/-0.03\\
16&13/06/2022(SB)&2.59/2.40&1.74&1.55&1.39&1.56&1.03/0.84\\
17&14/01/2023(SL)&3.14/3.09&1.33&1.19&1.02&1.18&1.96/1.91\\
18&10/07/2023(SB)&1.52/1.59&1.84&1.63&1.46&1.64&-0.12/-0.05\\
19&14/01/2024(SL)&1.72/1.82&1.51&1.35&1.19&1.35&0.37/0.47\\
20&29/01/2024(SL)&2.06/1.86&1.42&1.27&1.11&1.27&0.79/0.59\\
21&21/02/2024(SL)&2.98/3.11&1.56&1.40&1.24&1.40&1.58/1.71\\
22&14/05/2024(SB)&2.07/2.17&1.37&1.23&1.06&1.22&0.85/0.95\\
23&27/05/2024(SL)&2.29/2.42&1.94&1.71&1.53&1.73&0.56/0.69\\

 \hline
\end{tabular}
\\

\end{table}
\end{landscape}


In contrast, the black open circle indicates the estimated CME height at the onset time of the burst, obtained by extrapolation (panels a and c) and interpolation (panels b and d). The height of the Type II burst for the event at the burst onset time is indicated by a red triangle for comparison to the CME height. For the burst event of 14 Jan. 2024, the height of the Type II burst was estimated to be $\sim1.35 R_\odot$ at the onset time of the burst. It is evident from Fig.~\ref{Fig8}a, c that the $r_{TypeII}$ $<$ $r_{CME}$ for the burst event of 14 Jan. 2024. The height of the CME for the 21 Feb. 2024 burst event was also estimated using ellipsoid/GCS fits to the CMEs, as shown in Fig.~\ref{Fig8}(b, d). The red triangle in Fig.~\ref{Fig8}(b, d) marks the burst height, which is $\sim1.40 R_\odot$, and when it is compared to the CME height indicated by the black open circle, it is further evident that the CME height is more than the burst height for the Type II burst event on 21 Feb. 2024.

In this manner, we estimated the height of the remaining Type II burst events and the height of their corresponding CMEs, as listed in Table~\ref{tbl:3}. Comparing the height of CMEs and the bursts at the onset time of the burst, we estimated $\Delta$r = $r_{CME}$ - $r_{TypeII}$, as listed in the last column of Table~\ref{tbl:3}. Except for one event, in all the other 22 events, the value of $\Delta$r is positive. The positive values of the $\Delta$r for the 22 burst events indicate that the $r_{TypeII}$ $<$ $r_{CME}$.

\subsection{Quantitative estimate of anti-correlation between frequency bandwidth and CME angular width}
{{We combined coronal shock geometry with a coronal density model and plasma frequency relation to determine the frequency bandwidth of Type II bursts quantitatively and then correlated with the corresponding CME angular width, as explained in the following subsections.}}
\subsubsection{Shock surface reconstruction}
{We reconstructed a three-dimensional shock surface around a coronal shock using the method followed by \cite{su2022quantifying}, aiming to find out the spatial fraction of the shock surface having quasi-perpendicular geometry, which is a crucial parameter for the genesis of Type II radio emission. We assumed the coronal shock exhibits a 3-D bow shock geometry and constructed the shock surface using the following equation:}

\begin{equation}
z = h -  \frac{d}{s}\times {\bigg(\frac{\sqrt{R^{2}}}{d}\bigg)}^s
\end{equation}

{where h is the apex height of the shock, s is the bluntness of the shock representing the opening angle of the shock, and d is the semilatus rectum determining the width of the shock. The initial symmetric axis of the shock lies along the Z-axis, pointing from the CME's origin to its apex. In addition to these parameters, we further constrained two sets of angular parameters to reconstruct the shock surface. The first pair is the latitude and longitude ($\theta,\phi$) of the CME origin or the eruption location, while the other pair of latitude and longitude ($\theta',\phi'$) belongs to the apex or eruption direction.}                                                                \\

{Further, a rotation matrix is constructed to rotate the fitted shock structure from ($\theta,\phi$) to ($\theta',\phi'$):}

\begin{equation}
M_r =
\begin{bmatrix}
n_x n_x (1-\cos\alpha)+\cos\alpha &
n_x n_y (1-\cos\alpha)+n_z\sin\alpha &
n_x n_z (1-\cos\alpha)-n_y\sin\alpha \\

n_x n_y (1-\cos\alpha)-n_z\sin\alpha &
n_y n_y (1-\cos\alpha)+\cos\alpha &
n_y n_z (1-\cos\alpha)+n_x\sin\alpha \\

n_x n_z (1-\cos\alpha)+n_y\sin\alpha &
n_y n_z (1-\cos\alpha)-n_x\sin\alpha &
n_z n_z (1-\cos\alpha)+\cos\alpha
\end{bmatrix}
\end{equation}

where n is the rotation axis and taken as ${(e}$ $\times$ {e$'$)} and $\alpha$ is the rotation angle obtained by $cos^{-1}${(e.e$'$)}. The $n{_x}, n_{y},$ and $n_{z}$ represent the projection of the rotation axis {n} on the X-, Y-, and Z-axis, respectively. Taking all these parameters into account, we can now reconstruct the shock surface by adjusting them until they best match the CME-driven shock in SDO/AIA and STEREO/EUVI observations. In this manner, we fitted the shock surface for the three events, {\it{i.e.}} 08 March 2011, 25 October 2013, and 20 April 2014. The fitted shock surfaces for the events of 25 October 2013 and 20 April 2014 are shown in Fig~\ref{Fig10}(a, b) respectively. The blue isolines overplotted on the CME structures in SDO/AIA 193{\AA} represent the 3D fitted shock surface.
%
\begin{figure} [!h]
   \centerline{\hspace*{0.04\textwidth} 
               \includegraphics[width=1.0\textwidth,clip=]{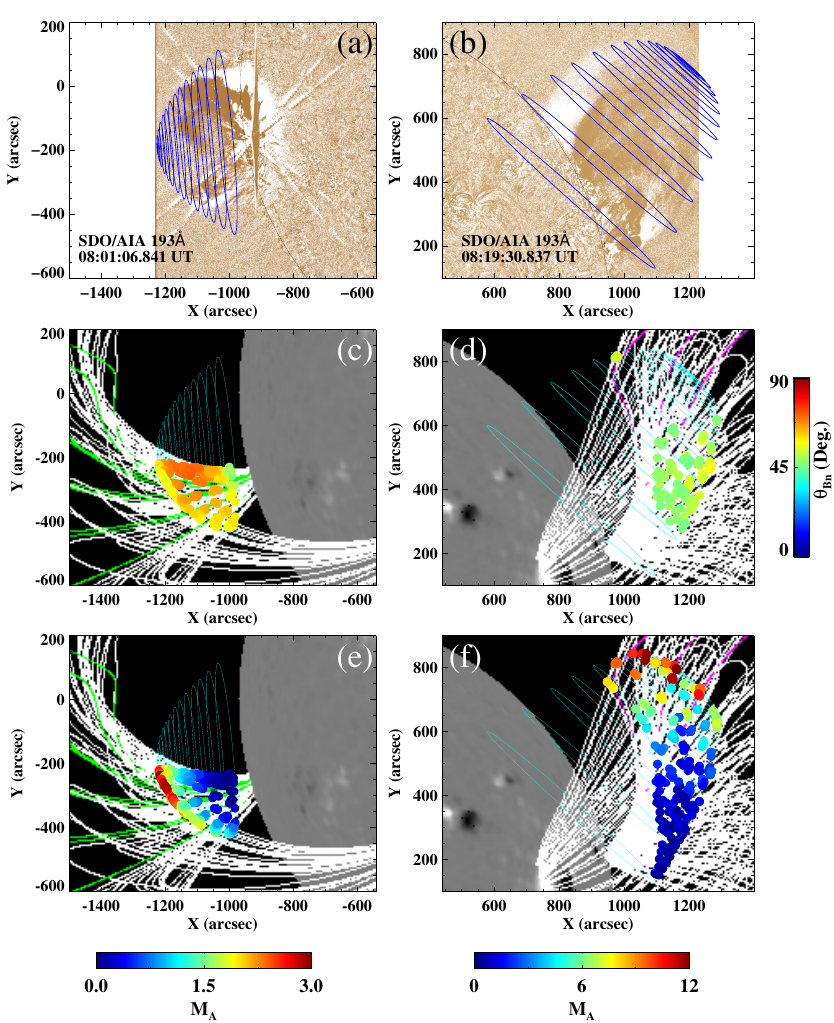}
               \hspace*{0.06\textwidth}
              }
     \vspace{-0.36\textwidth}   
     \centerline{\Large \bf     
      \hspace{0.0 \textwidth}
      \hspace{0.415\textwidth}
         \hfill}
     \vspace{0.32\textwidth}    

\caption{(a, b) Running difference image of the Sun from SDO/AIA 193{\AA} showing eruption of the CME associated with the 25 October 2013 and 20 April 2014 Type II radio burst, respectively. The blue isolines are the 3D fitted surface reconstruction of the shock. (c, d) Selected magnetic field lines extrapolated using PFSS around the shock surface. Green/pink and white lines show open and closed field lines, respectively. The cyan isolines are reconstructed shock surfaces. The distribution of $\theta_{Bn}$ $> 45^\circ$ on the shock surface is depicted in multiple colors, while the values of $\theta_{Bn}$ are represented by a colorbar on the right of the panels. (e, f) Same as Fig. (c, d), but the distribution of Mach number is overlaid on the shock surface, with the corresponding Mach number values represented by the bottom horizontal colorbars.}
\label{Fig10}
\end{figure}
%
\subsubsection{Quasi-perpendicular region of the shock surface}
{To obtain the quasi-perpendicular region of the shock surface, we need the magnetic field around the shock surface. In the present study, Type II bursts are produced by shock-accelerated electrons in the upstream region of the shock surface. Therefore, we used a potential-field source-surface (PFSS) model \citep{schrijver2003photospheric} to extrapolate the magnetic field around the shock front. The magnetic field lines obtained using a PFSS model in the upstream region of the shock are not disturbed by the shock, and hence we preferred to use the PFSS model. Figure ~\ref{Fig10}(c, d, e, and f) shows the extrapolated magnetic field lines obtained from the PFSS model. The white lines shown in Fig. ~\ref{Fig10}(c, d, e, and f) are the closed magnetic field lines, while the green and pink lines represent the open magnetic field lines. To determine the quasi-perpendicular condition at the shock front, we selected only the field lines surrounding it and then estimated the angle between the tangent (t) of the magnetic field line and the shock normal (n), represented as ($\theta_{Bn}$). The angle $\theta_{Bn}$ between {t} and {n} can be calculated as $cos^{-1}${(t.n)}. We estimated the $\theta_{Bn}$ for those field lines which are close ($<$ 0.015 $R_{\odot}$) to the shock front and lie in the upstream region of the shock surface. }
{As mentioned earlier, the PFSS model is useful for estimating the $\theta_{Bn}$ in the upstream region. 
The region on the shock surface with $\theta_{Bn}$ less than $45^\circ$ are known as the quasi-parallel region, while the region with $\theta_{Bn}$ more than $45^\circ$ are known as the quasi-perpendicular region. Figure~\ref{Fig10}(c, d) shows the distribution of the $\theta_{Bn}$ on the shock surface for the events of 25 October 2013 and 20 April 2014, respectively. Since we estimated the $\theta_{Bn}$ for those field lines that are close enough to the shock surface and are in the upstream region, the distribution of $\theta_{Bn}$ on the shock front for both events shows that the quasi-perpendicular region is restricted to the specific regions on the shock surface. It is clear from Figure~\ref{Fig10}(c, d) that with the increase in angular width of the CME, the region satisfying the quasi-perpendicular condition decreases.}\\

\subsubsection{Alfv\'en Mach-Number of the shock surface}

{Alfv\'en Mach Number$(M_{A})$ along with $\theta_{Bn}$ is the key parameter which determines the efficient acceleration of the electrons along the shock surface. The Alfv\'en Mach number is the ratio of the shock speed($v_s$) with the Alfv\'en speed ($v_{A}$), where $v_A$ is given as follow:}

\begin{equation}
v_A = \frac{B}{\sqrt{\mu_0 n}}
\end{equation}

{The above equation can be written as $v_A$ = $2.8\times10^{16}(B/\sqrt n)$, where B is the magnetic field in Tesla, n is the density in $m^{-3}$, and $v_A$ is the Alfv\'en speed in m/s \citep{priest2014magnetohydrodynamics}. The magnetic field is obtained from the PFSS model for the field lines around the fitted shock surface. PFSS provides the 3D magnetic field in terms of radial, longitudinal, and latitudinal components represented by $B_r$, $B_{\theta}$, and $B_{\phi}$, respectively. The total magnetic field B is given by $\sqrt{B_r^{2}+B_\theta^{2}+B_\phi^{2}}$. n is the electron density obtained using the Newkirk density model. Further, to estimate the local shock speed at each mesh point of the shock surface, we reconstructed the shock surface at two time stamps separated by a time interval $(\Delta t)$. Both the shock surfaces were reconstructed using independent adjustment of the shock parameters. After shock reconstruction, we estimated the displacement between each mesh point of the fitted shock surface. For each mesh element, the displacement was projected onto the local shock normal to obtain the normal propagation distance($\Delta S_n$). The local shock speed at each mesh point is then estimated as:}

\begin{equation}
v_s = \frac{\Delta S_n}{ \Delta t}
\end{equation}

{Figure~\ref{Fig10}(e, f) shows the distribution of the $(M_{A})$  on the shock surface for the events of 25 October 2013 and 20 April 2014, respectively. It is clear from Figure~\ref{Fig10}(e, f) that with the increase in angular width of the CME, the region satisfying the $(M_{A})$ $>$ 1.5 doesn't change significantly.}

\subsection{Estimation of frequency bandwidth}
{Once we determined the quasi-perpendicular shock region and Mach number condition, we obtained the maximum height ($h_{max}$) and the minimum height ($h_{min}$) along the shock surface. Further, combined with a 1-fold Newkirk density model, we converted the height to density, then used the plasma frequency relation to convert the density to the frequency. Finally, we estimated the frequency bandwidth ($\Delta$f = $f_{max} - f_{min}$),  where $f_{max}$, $f_{min}$ represent the frequency corresponding to $h_{min}$ and $h_{max}$, respectively.} {Table~\ref{tbl:4} lists the frequency bandwidth obtained in the above manner for three events on 08 March 2011, 25 October 2013, and 20 April 2014, for which both SDO/AIA and STEREO/EUVI observations were available. It is evident from Table~\ref{tbl:4} that the event on 25 October 2013 with a minimum angular width of 25.41$^\circ$ has the maximum bandwidth of 79.62 MHz, while the other two events on 20 April 2014 and 8 March 2011 with high angular widths of 33.58$^\circ$ and 54.64$^\circ$ respectively, have a lesser frequency bandwidth of 52.73 and 60.39 MHz, respectively.}

\begin{table}
 \caption{Quantitative estimation of frequency bandwidth}
 \label{tbl:4}
\begin{tabular}{cccc}
\hline
 \multicolumn{1}{c}{}
Serial& Event & Angular & Frquency \\
No.&    Date  & Width (Deg.)  &Bandwidth (MHz)\\
\hline

1&25/10/2013&25.41 &79.62\\
2&20/04/2014&33.58	&52.73\\
3& 08/03/2011 & 54.64&60.39\\
 \hline
 \end{tabular}
 \end{table}

\section{Summary and Discussion}
\label{discuss}
In the present study, we estimated the frequency bandwidth of 23 metric Type II burst events. The estimated frequency bandwidth was correlated with the angular width of the associated CMEs, as determined using a GCS forward-fitting model. Our analysis revealed an anti-correlation (cc =- 0.62) between the frequency bandwidth and the instantaneous angular width. The obtained correlations suggest that the frequency bandwidth of the metric Type II bursts is anti-correlated with the angular width of the associated CMEs, implying coronal Type II bursts associated with CMEs could be produced in a narrower region. Further, in our study, when comparing the heights of CMEs and Type II bursts, we found that the heights of Type II bursts for all (except one) were lower than those of the associated CMEs. It was suggested earlier in their studies \citep{gopalswamy2009relation,majumdar2021imaging} that if the height of a CME exceeded that of a Type II burst, the Type II burst could be located in the CME's flank region. Hence, we suggest that Type II bursts may be produced in the flank region of CMEs rather than at the CME front. Our results contrast with those of the study by {\cite{ramesh2022new}}, which found that Type II emission associated with CMEs of large angular extent has a broader frequency bandwidth.
\begin{figure} [!h]   
   \centerline{\hspace*{0.04\textwidth} 
               \includegraphics[width=\textwidth,clip=]{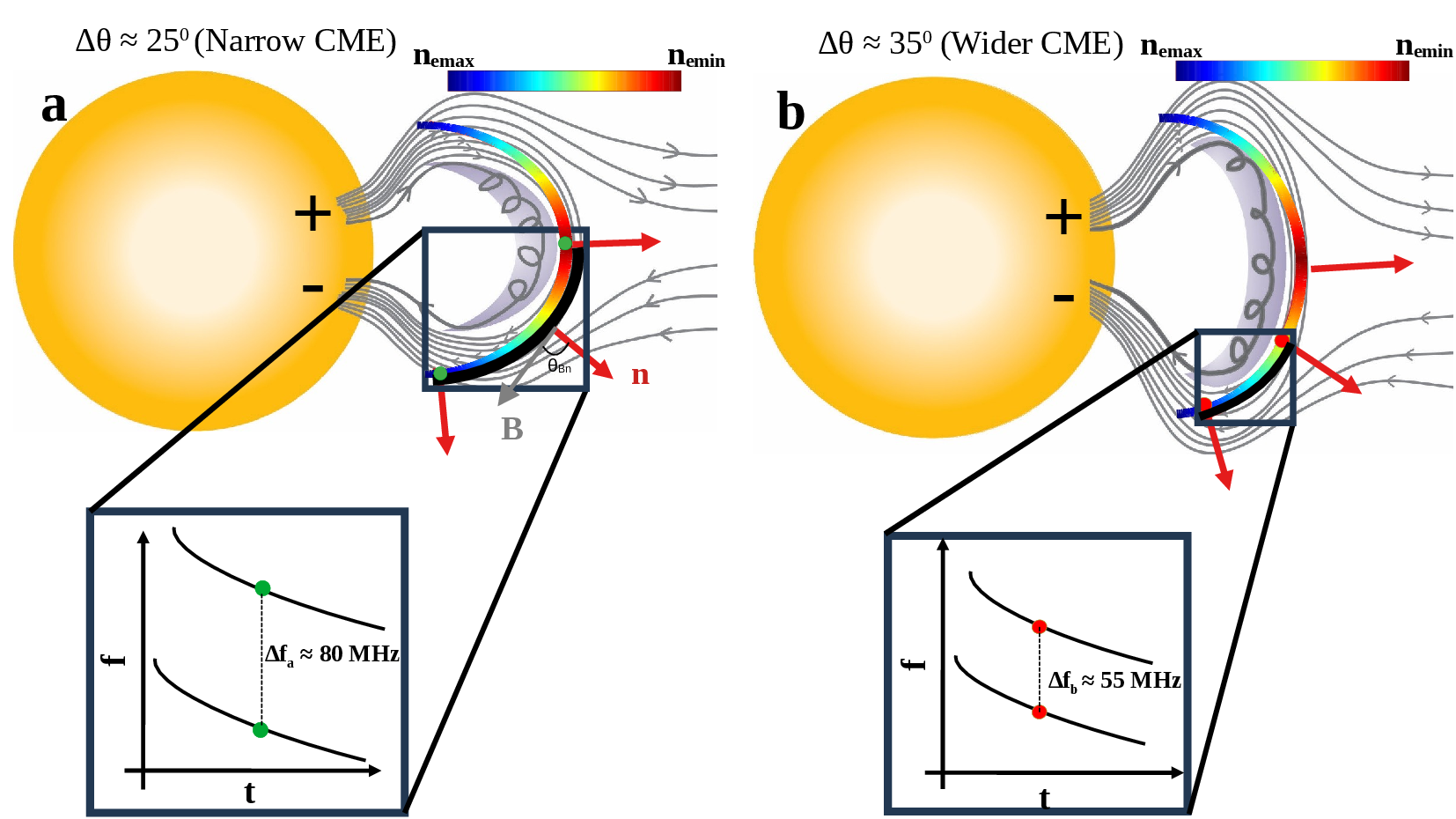}
               \hspace*{0.06\textwidth}
              }
     \vspace{-0.36\textwidth}   
     \centerline{\Large \bf     
      \hspace{0.0 \textwidth}
      \hspace{0.415\textwidth}
         \hfill}
     \vspace{0.32\textwidth}    

\caption{a) A schematic representation for illustrating the genesis of Type II emission of broader frequency bandwidth associated with the CME of a smaller instantaneous angular width. The insert plot is a frequency-time plot showing a broader frequency bandwidth. b) A schematic representation for illustrating the genesis of Type II emission of a narrower frequency bandwidth associated with the CME of a larger instantaneous angular width. The insert plot is a frequency-time plot showing a narrower frequency bandwidth. {The thick color gradient curve in both panels is the CME-driven shock front. The black-filled curve on the flank of the CME marks Type II emission produced by shock-accelerated electrons.}}
   \label{Fig11}
   \end{figure}

Generally, CMEs with a larger angular extent will produce a broader shock with different locations of the shock, from the nose to the flanks, at different heliocentric distances. Thus, for a larger CME, a wide range of plasma density variations will be noticed from the nose to the flank region of the CME. Hence, a broad range of frequencies for the resulting Type II emissions is usually expected, consistent with the results reported by {\cite{ramesh2022new}}. 
Our results, however, are the opposite of the aforementioned scenario and suggest that a CME with a larger angular extent can give rise to Type II emission with a narrower frequency bandwidth. {We carried out a quantitative estimate of the frequency bandwidth, based on the quasi-perpendicular shock geometry and a higher-Mach-number condition, for three Type II burst events with different angular widths. It was found that, for the CME with the smaller angular width, the quasi-perpendicular shock geometry and a high Alfv\'en Mach Number were satisfied over a large region of the shock front, from the nose to the shock flank, producing a Type II burst with a broader frequency bandwidth. On the other hand, for the CME with the larger angular width, the quasi-perpendicular shock geometry and high Alfv\'en Mach Number were satisfied in a relatively narrower region of the shock front, resulting in a Type II burst with a narrower frequency bandwidth.}  A schematic, as shown in Fig.~\ref{Fig11}, depicts the scenario of Type II emissions of broader or narrower frequency bandwidth in the flank regions of the associated CMEs. Figure \ref{Fig11}(a) shows a CME of smaller instantaneous angular width. The resulting Type II radio emission due to shock-accelerated electrons is produced in a broader region shown by a black curve on the flank of the CME, where high Mach number and quasi-perpendicular conditions are being satisfied. The insert in Fig.~\ref{Fig11}(a) shows a frequency-time plot with a broader frequency bandwidth for the Type II burst associated with the CME of a smaller angular width. On the other hand, Fig.~\ref{Fig11}(b) shows a CME of larger instantaneous angular width. Similar to the earlier case, the resulting Type II radio emission in this case is produced in a narrower region than the earlier case, shown by a black curve on the flank of the CME, where the high Mach number and quasi-perpendicular conditions are being satisfied. The insert in Fig.~\ref{Fig11}(b) shows a frequency-time plot with a narrower frequency bandwidth for the Type II burst associated with the expanding CME of larger angular width.

It was discussed by {\citep{knock2005type}} that a large and faster CME could be radio-poor if the quasi-perpendicular shock geometry were not satisfied or a sufficient number of electrons was not accelerated. In addition to the shock geometry, for radio emissions to occur, the shock speed must exceed the Alfv\'en speed. Considering these favorable conditions, our results show that the conditions for efficient electron acceleration are not satisfied in all regions of the CME. Only a certain region may favor conditions for electron acceleration; in that case, the radio emissions can result from a narrow region of the CME. Therefore, it may happen that despite having a large angular width, a CME will produce narrow-band emission.

{A radially propagating shock will be quasi-parallel if the magnetic field lines are oriented along the radial direction,} and quasi-parallel shocks are not considered efficient producers of Type II emission. However, they can efficiently produce Type II emissions from their laterally expanding quasi-perpendicular flank regions. In the case of lateral expansion of the shock, we can observe broadband emission because the shock's flank encounters a large coronal plasma region with varying densities. Therefore, we can observe a broad range of frequencies for Type II radio emissions {\citep{knock2005type}}. Even then, the flank regions of CMEs can produce narrow-band Type II emission due to frequency blocking. {\cite{knock2003type}} mentioned that emission may not propagate through the shock when it is located in a high-density plasma region. In such cases, the observer can not detect the emission through some parts of the shock front. Hence, they will see emission from a restricted region of the shock front, eventually leading to narrow-band emissions. For a radially propagating shock, when the flank region is not responsible for the radio emission, a small, localized portion of the shock accelerates electrons, producing narrow-band radio emission.

In the present study, we select the fundamental branch of the Type II emission for the statistical correlation of the frequency bandwidth with the angular width. It is known that in metric Type II bursts, the harmonic emission from the shock is usually stronger than the fundamental. Therefore, we will investigate the harmonic Type II emission in a subsequent study and verify whether an anti-correlation exists between the frequency bandwidth and the angular width for the harmonic emission. Also, it will be intriguing to investigate how the correlation varies when the study is expanded to DH-Type II bursts. 

\begin{acks}
This work used the open data catalogs from GLOSS at the Gauribidanur Observatory, CULGOORA, and e-Callisto. We acknowledge the STEREO/SECCHI, SDO/AIA, and SOHO/LASCO teams for making their data publicly available.{PFSS code developed by C.J. Schrijver and M.L. DeRosa, available under SolarSoft IDL, was used for magnetic field extrapolation.} We thank the anonymous reviewer for his constructive comments and suggestions, which helped us improve the paper. 
\end{acks}

\begin{authorcontribution}
MR performed all the analyses and wrote the initial manuscript. SKB conceived the initial research concept, edited the manuscript, and submitted it. JP helped in editing the manuscript. All authors contributed to the writing of the manuscript.
\end{authorcontribution}




\begin{dataavailability} 
The data for solar radio bursts is taken from Gauribidanur (GLOSS)(\url{https://www.iiap.res.in/solarradioimages\#spectrogram}), CULGOORA(\url{https://www.sws.bom.gov.au/World\_Data\_Centre/1/9}), and e-CALLISTO(\url{https://www.e-callisto.org/Data/data.html}). The CME data used here is obtained from SOHO/LASCO (\url{https://cdaw.gsfc.nasa.gov/CME_list/}), STEREO (SECCHI)(\url{https://vso.nascom.nasa.gov/cgi/search?time=1&instrument=1&spectrum=1&version=current&build=1}), and SDO (AIA)(\url{http://jsoc.stanford.edu/ajax/lookdata.html}).
\end{dataavailability}



\begin{ethics}
\begin{conflict}
The authors declare that they have no conflicts of interest.
\end{conflict}
\end{ethics}

\bibliographystyle{spr-mp-sola}
\bibliography{References}  

\IfFileExists{\jobname.bbl}{} {\typeout{}
\typeout{****************************************************}
\typeout{****************************************************}
\typeout{** Please run "bibtex \jobname" to obtain} \typeout{**
the bibliography and then re-run LaTeX} \typeout{** twice to fix
the references !}
\typeout{****************************************************}
\typeout{****************************************************}
\typeout{}}

\end{document}